\documentclass[format=sigconf]{acmart}

\usepackage{tabularx}
\usepackage{subcaption}
\usepackage{csvsimple}
\usepackage{filecontents}

\newcolumntype{Y}{>{\centering\arraybackslash}X}

\begin{document}
\title[]{A new encoding of implied volatility surfaces for their synthetic generation}

\author{Zheng Gong}
\affiliation{%
	\institution{University of Essex}
	\city{Colchester}
	\postcode{CO4 3SQ}
	\country{United Kingdom}
}
\additionalaffiliation{%
	\institution{UBS}
	\city{London}
	\postcode{EC2M 2QS}
	\country{United Kingdom}
}
\email{zg19500@essex.ac.uk}

\author{Wojciech Frys}
\affiliation{%
	\institution{UBS}
	\city{London}
	\postcode{EC2M 2QS}
	\country{United Kingdom}
}
\email{wojciech.frys@ubs.com}

\author{Renzo Tiranti}
\affiliation{%
	\institution{UBS}
	\city{London}
	\postcode{EC2M 2QS}
	\country{United Kingdom}
}
\email{renzo.tiranti@ubs.com}

\author{Carmine Ventre}
\affiliation{%
	\institution{King's College London}	
	\city{London}
	\postcode{WC2R 2LS}
	\country{United Kingdom}
}
\email{carmine.ventre@kcl.ac.uk}

\author{John O'Hara}
\affiliation{%
	\institution{University of Essex}
	\city{Colchester}
	\postcode{CO4 3SQ}
	\country{United Kingdom}
}
\email{johara@essex.ac.uk}

\author{Yingbo Bai}
\affiliation{%
	\institution{UBS}
	\city{London}
	\postcode{EC2M 2QS}
	\country{United Kingdom}
}
\email{yingbo.bai@ubs.com}


\begin{abstract}
In financial terms, an implied volatility surface can be described by its term structure, its skewness and its overall volatility level. We use a PCA variational auto-encoder model to perfectly represent these descriptors into a latent space of three dimensions. Our new encoding brings significant benefits for synthetic surface generation, in that (i) scenario generation is more interpretable; (ii) volatility extrapolation achieve better accuracy; and, (iii) we propose a solution to infer implied volatility surfaces of a stock from an index to which it belongs directly by modelling their relationship on the latent space of the encoding. All these applications, and the latter in particular, have the potential to improve risk management of financial derivatives whenever data is scarce. 
\end{abstract}

\maketitle

\begin{figure*}[!t]
\centering
\caption{Variational Auto-Encoder Model}
\label{fig:vae}
\includegraphics[width=0.8\textwidth]{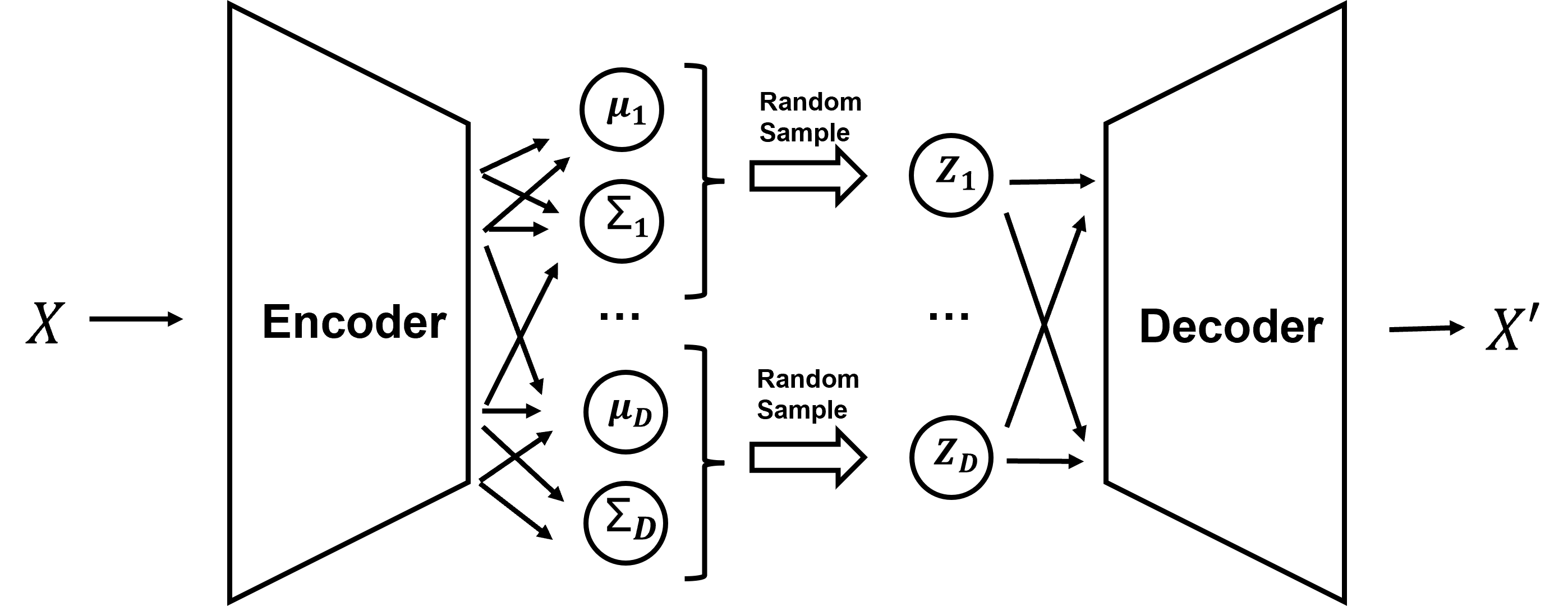}
\end{figure*}

\section{Introduction} 
\label{sec:intro}
The Black-Scholes-Merton (BSM) model \cite{black2019pricing,merton1973theory} is the benchmark for stock option pricing and valuation. One major weakness of the standard BSM method is the assumption that volatility of the underlying stock is a constant, and not related to the moneyness or the remaining term of the option contract.  This does not conform with our observations in the market. 

If we equate the BSM formula with market option prices and solve for the volatility parameter, we get the \emph{implied volatility} value. Implied volatilities have a typical characteristic, which is, the value of volatility gets smaller as the option moneyness move close to one and gets larger as it moves away from one. This phenomenon is referred to as the implied Volatility Smile or Volatility Skewness \cite{andersen2000jump}. If we take the remaining term of an option contract as another dimension into the volatility smile, we get the Implied Volatility Surface. 

Implied volatility surface is the most important variable to consider if anyone wants to design, trade or evaluate financial derivatives. 
The most straightforward use case is to calculate the option prices of non-existent term and moneyness combinations of a particular underlying stock \cite{homescu2011implied} on a particular trading day. This involves the process of predicting an extended subset of volatility values on a surface from some existing values that can be observed from the market. This task is commonly referred to as the implied volatility extrapolation. Once the extended volatility values are extrapolated, they can be put into the BSM formula to calculate the required prices of financial derivatives of the special term and moneyness. The most common scenario is to extrapolate from products with short-term expiration to products with long-term expiration. The profit or loss of issuing such products are largely depended on the accuracy of the extrapolated implied volatility values.

Moving one step further, we could not only predict the extended term and moneyness combinations, but also predict implied volatility surfaces of a specific stock from another related stocks or stock index. We consider a scenario where, on a particular trading day, we have zero information about a stock's implied volatility surface, and therefore we cannot perform extrapolation. However, we have the historical implied volatility surfaces of this stock and some other related stocks or stock indexes. We need to identify the relationships between different stocks/indexes from their empirical volatility surfaces and predict future implied volatility surfaces from scratch. This task is most challenging, and to the best of our knowledge, there is no established method to solve this problem.

It is also believed that the shape of an implied volatility surface reflects the current market perception of risk and return, and the implied volatility values indicate the demand and supply for different combinations of moneyness and term structure \cite{aijo2008implied}. Ideally, market participants could generate many implied volatility surfaces with similar shape but not identical values, all representing typical market environments, and use this batch of synthetic surfaces 
for evaluating their derivative trading engines. Because stress scenarios are rarely represented in the historical data, the ability to generate synthetic implied volatility surfaces of good quality and in an interpretable manner is crucial. 

In this paper, we present a new encoding mechanism of implied volatility surfaces using a PCA variational auto-encoder model. The new encoding significantly improves the state of the art 
for the three problems mentioned above.

There is limited literature on the modelling of implied volatility surfaces with neural network approaches \cite{cao2020neural,ackerer2020deep,bergeron2022variational,zheng2019gated}. Their work did not clarify the question of what features of the implied volatility surfaces are exactly encoded or learnt by the neural network model. This is a crucial issue, as we want to generate (a subset or the whole of) a volatility surface to perform the aforementioned task. If we do not know how the neural network model encodes the market data or what is represented in the neural network model, we cannot interpret the generated synthetic surfaces

The remainder of the paper is organised as follows. Section \ref{sec:review} introduces background knowledge about stock options, and briefly discusses the relevant literature on auto-encoder neural network models for implied volatility surfaces. In that section, we also explain our PCA variational auto-encoder model, which is used to encode the implied volatility surface differently. Section \ref{sec:data} introduces the data we used for our experiments and, more importantly, how we evaluate the usefulness of our generated synthetic implied volatility surfaces. Section \ref{sec:encoding} introduces our method of encoding implied volatility surfaces; we compare our approach with an alternative solution highlighting the differences on the training process, as well as the unique characteristics of the encoded latent space. Section \ref{sec:generate} shows how we can leverage the new encoding to efficiently solve the three challenges on synthetic surfaces generation; interpretable scenario-based generation, implied volatility extrapolation and stock-specific generation. Section \ref{sec:conclusion} concludes our work and proposes directions for future research.

\section{Literature Review}
\label{sec:review}
\subsection{Stock Option and Implied Volatility Surface}
A European stock option gives an investor the right, but not the obligation, to buy or sell a stock at a predefined strike price $K$ and expiration date $T$. The stock specified in the option contract is called the underlying stock. At a date $t$ during the contract period, the price of underlying stock is $S_{t}$, and the remaining term of the option contract is $\tau = T - t$. The ratio between the strike price and price of the underlying asset is called moneyness $M_{t} = \frac{K}{S_{t}}$.

Stock options are financial derivatives, and at the time $t$, their prices $P_{t}$ are usually calculated from BSM formula \cite{davis2010black}:  
\begin{equation*} 
\label{eq:bsm}
P_{t} = BSM(S_{t}, K, T, r, t, \sigma_{t}),
\end{equation*}
where $r$ is the risk-free interest rate, and we assume there are no dividends to be paid from the underlying stock.

The $\sigma_{t}$ is the key parameters in the BSM formula. As we mentioned in Section \ref{sec:intro}, if we solve the BSM function using $P_{t}$ observed from the market, we obtain the value of $\sigma_{t}$ which is named implied volatility, and it is actually a function of $\tau$ and $M_{t}$. The function $\sigma_{t} = f(\tau, M_{t})$ is usually represented as the implied volatility surface.

\subsection{Variational Auto-Encoder and Implied Volatility Surface}
Auto-encoder (AE) is a typical structure for neural network models, and has been very popular in the areas of image and language processing \cite{zhang2018better,chen2017kate}. It is effectively using two separate neural networks (encoder and decoder) to approximate the identify function $f(X) = X$, by first encoding the input data $X$ into a latent vector $Z$ and then decoding $Z$ to output $X'$ that replicate the input data $X$. When the difference between $X$ and $X'$ is small enough, the latent representation $Z$ is considered to have successfully stored the information about the input $X$ in the latent space, and usually the dimensionality of $Z$ is much smaller than $X$ (i.e., dimensionality reduction). 

Variational Auto-Encoder (VAE) models are a popular extension of the autoencoder framework. It was first proposed by 
\cite{kingma2013auto}, and 
later expanded 
in finer details \cite{kingma2019introduction}. A variational auto-encoder model structure is shown in Figure \ref{fig:vae}. 

Instead of producing latent encodings $Z$ in a deterministic manner like the classic auto-encoder, the VAE randomly samples them from a normal distribution that is parameterized by the encoder neural network. This special arrangement has one particular advantage. 
Because encoder and decoder networks are not directly linked, when the encoder takes the same input batch in training, the changes in $Z$ are not only due to the updated parameters in the encoder, but also the randomness caused by the sampling process. This forces the decoder network to learn a distribution of $Z$ values instead of deterministic values, which results in a more powerful decoder for synthetic data generation. This also 
explains why a VAE model 
takes longer training time than a classic AE model on the same dataset, even with identical training parameters. 

The loss function for training a VAE model is formulated as: 
\begin{equation*} 
\label{eq:vae}
L = L_{recon} + \lambda_{kl}\cdot L_{kl},
\end{equation*}
where $L_{recon}$ is the reconstruction error, and it is calculated as the mean absolute error (MAE) between input $X$ and output $X'$; $\lambda_{kl}$ is the weight of $L_{kl}$ in the loss function; and, $L_{kl}$ is the Kullback-Leibler (KL) divergence between the encoded distributions and standard normal distribution $N(0,1)$:
\begin{equation*}
\label{eq:lkld}
L_{kl} = -\frac{1}{2} \cdot \sum_{k=1}^{D}(1 + \log\Sigma_{k}^{2} - \Sigma_{k}^{2} - \mu_{k}^{2})
\end{equation*}
$D$ being the number of latent dimensions, $\Sigma_{k}$ and $\mu_{k}$ produced by the encoder neural network as shown in Figure \ref{fig:vae}. It is important to note that $\lambda_{kl}$ is a hyperparameter that tunes the strength of the regularization provided by KL. Naturally, when  $\lambda_{kl}$ approaches 0, then the learning process optimizes $\Sigma_{k}$ to be also close to 0 and the VAE effectively becomes a classic AE.

Bergeron et al. \cite{bergeron2022variational} implemented variational auto-encoder models on the implied volatility surfaces from the foreign exchange market. They encoded implied volatility surfaces into various numbers of latent dimensions, but did not clarify how many dimensions are optimal. This is due to the fact that their method could not separate the features of a surface into each latent dimensions, so that every time a new dimension is added, the model gains a certain degree of accuracy. Whether that dimension is really encoding a meaningful feature or just market noise is however unanswered. In addition to that, the encoding vectors obtained from their model are not useful for stock specific generation due to the lack of interpretability.

\subsection{PCA Variational Auto-Encoder}
To address the problems discussed above, we upgrade the variational auto-encoder model to a PCA variational auto-encoder. 

The PCA auto-encoder model was first developed by 
\cite{ladjal2019pca}. They made an important change to the loss function of a classic auto-encoder by adding an extra term $L_{cov}$, which measures the covariance between all pairs of latent dimensions. They try to reduce the value of $L_{cov}$, which effectively forces the latent dimensions to be independent to each other. The formula for $L_{cov}$ is: 
\begin{equation} 
\label{eq:lcov}
L_{cov} = \sum_{i=1}^{D}\sum_{j=i+1}^{D} \left[ \frac{1}{B}\sum_{t=1}^{B}(Z_{t,i}Z_{t,j}) - \frac{1}{B^2}\sum_{t=1}^{B}Z_{t,i}\sum_{t=1}^{B}Z_{t,j}\right]
\end{equation}
where $B$ is the number of data point in a training batch, and $Z_{t,i}$ is the encoding of $t$th data sample for $i$th dimension. It is imperative that the encoded value is from the same random sampling as used in the calculation of $L_{recon}$, not the $\mu_{i}$ used in $L_{kl}$. We performed ten trainings using both encoding versions. In the versions that used $\mu_{i}$, only one dimension was found to be significant (as described in Section \ref{ssec:training-PCA-VAE}) in all ten cases, however the version that use $Z_{t,i}$ has found all 3 significant dimensions every time.

We combine the concept of PCA auto-encoder with the variational auto-encoder models. The final loss function for training our PCA variational auto-encoder becomes:
\begin{equation} 
\label{eq:pca}
L = L_{recon} + \lambda_{kl} \cdot L_{kl} + \lambda_{cov} \cdot L_{cov}.
\end{equation}
The value of $\lambda_{cov}$ controls how much restriction we want to enforce on the model due to the covariance. On one hand, if $\lambda_{cov}$ is too small, the latent dimensions are not independent. On the other hand, the model loses the ability to generate surfaces if $\lambda_{cov}$ is too large. By trial and error, we found the optimal value of $\lambda_{cov}$ is $0.1$ for training a PCA variational auto-encoder on implied volatility surfaces.

We will be using the VAE model structure together with this loss function to train the PCA variational auto-encoder models. We are going to demonstrate their unique characteristics for the model training process and encoded latent space, and their strengths for dealing with the three synthetic surface generation challenges mentioned in the introduction. 

\section{Dataset and Evaluation Criteria}
\label{sec:data}
Our dataset consists of daily implied volatility surfaces for the period from 4 October 2016 to 4 October 2021. We collected data from 44 European listed stocks, as well as the stock index of STOXX50\footnote{STOXX50 is a stock index covers 50 stocks from the Eurozone}. We also have the daily high, low, open, close prices of these stocks and index over the same period of time. The volatility surfaces computed before 4 October 2020 are marked as training datasets, and those after that date are testing datasets.

Among the 44 stocks, we take 6 stocks out of the dataset for performing the volatility surface extrapolation and specific stock generation tasks. We want to emphasize that the PCA variational auto-encoder model is capable of learning the general features of volatility surfaces, which not only exist among the temporal dimensions but also between difference stocks.

After we generate synthetic implied volatility surfaces from our model, we wish to measure how realistic or useful they are. To evaluate a model, one often calculates the numerical difference, as, e.g., the mean absolute error (MAE), between every point on a real surface and the generated surface. The conclusion would be that if MAE values are smaller, the model performs better. However, this seems unsatisfactory for two reasons. Firstly, the financial dataset sampled from the market contains a lot of noise. Consequently, we should not force the model to learn every piece of the data and the aforementioned evaluation method would more easily lead to overfitting. 
The MAE and other similar measurements are more adequate for computer science tasks, where the data has much less misleading information. Secondly, if we generated a surface that is identical to the market surface (i.e., with zero MAE) then it would not be useful in our context. We want our synthetic surfaces to be similar but not identical to the market, and the similarity is within a reasonable range (e.g. bid/offer spread), so that the synthetic surfaces are considered useful. In order to obtain a measure of usefulness, we collected 9082 market quotes for options on European stocks, and calculated the implied volatility spread between bid quotes and offer quotes, and we come up with the evaluation threshold as listed in Table \ref{tab:evaluation_threshold}. The table needs to be read in the following way. If, for example, the absolute difference between the market implied volatility and generated implied volatility $|\sigma_{3,0.9} - \sigma'_{3,0.9}|$ is smaller than $0.0149$, then we consider the generated $\sigma'_{3,0.9}$ as a satisfactory implied volatility point. We calculate the percentage of satisfactory points within all the points on all the surfaces we generate, and call this the \emph{satisfaction rate}. In other words, if the generated implied volatility is within the range of bid offer spread from the market value, we consider it as satisfactory.  
 
\begin{table}
\centering
\caption{Evaluation Thresholds}
\label{tab:evaluation_threshold}
\begin{tabular}{cccc} 
\toprule 
& $M\in(0,0.9]$ & $M\in(0.9,1.05]$ & $M\in(1.05,\infty)$  \\
$\tau\in(0,3]$ & 1.49\% & 1.83\% & 1.69\% \\
$\tau\in(3,9]$ & 0.88\% & 1.18\% & 1.05\% \\
$\tau\in(9,\infty)$ & 0.90\% & 0.98\% & 1.09\% \\ 
\bottomrule
\end{tabular}
\end{table}

\begin{figure}
  \centering
  \subcaptionbox{Classical Variational AutoEncoder with 3 Latent Dimensions\label{fig:classicialtraining}}{\includegraphics[width=0.9\columnwidth
  ]{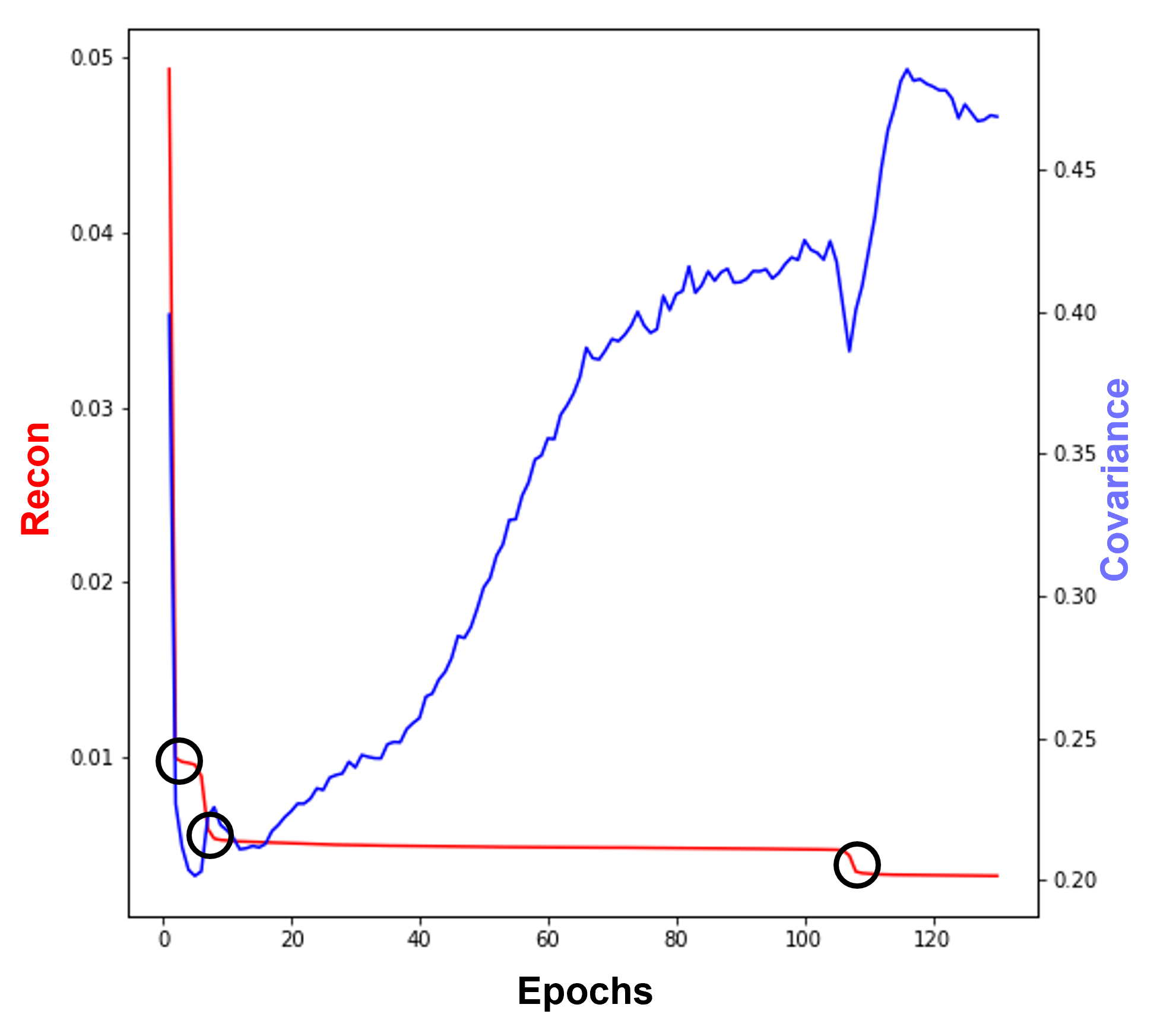}}\hfill%
  \subcaptionbox{PCA Variational AutoEncoder with 3 Latent Dimensions \label{fig:PCAtraining}}{\includegraphics[width=0.9\columnwidth
  ]{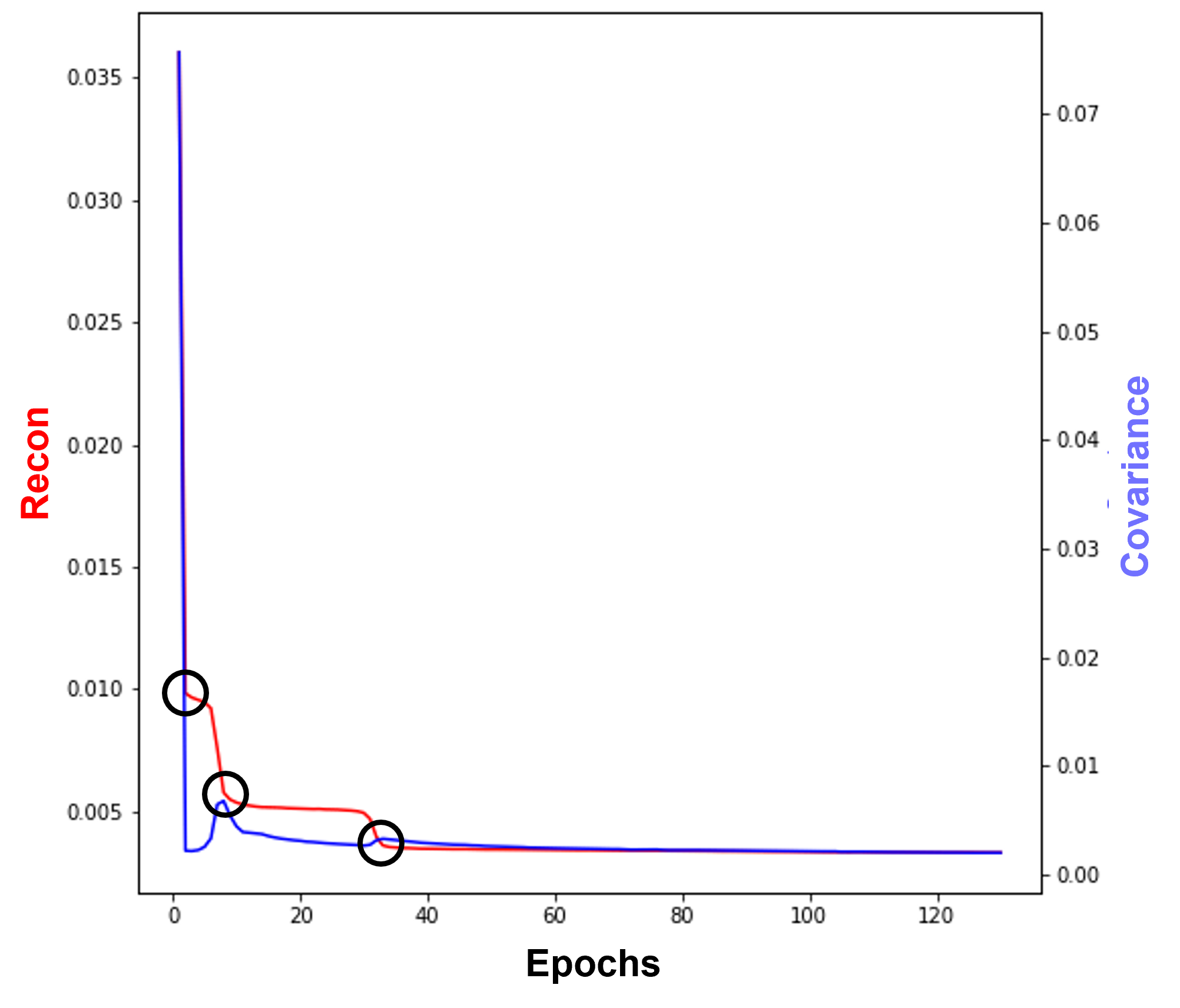}}\hfill%
  \subcaptionbox{PCA Variational AutoEncoder with 4 Latent Dimensions \label{fig:PCAtraininglarge}}{\includegraphics[width=0.9\columnwidth
  ]{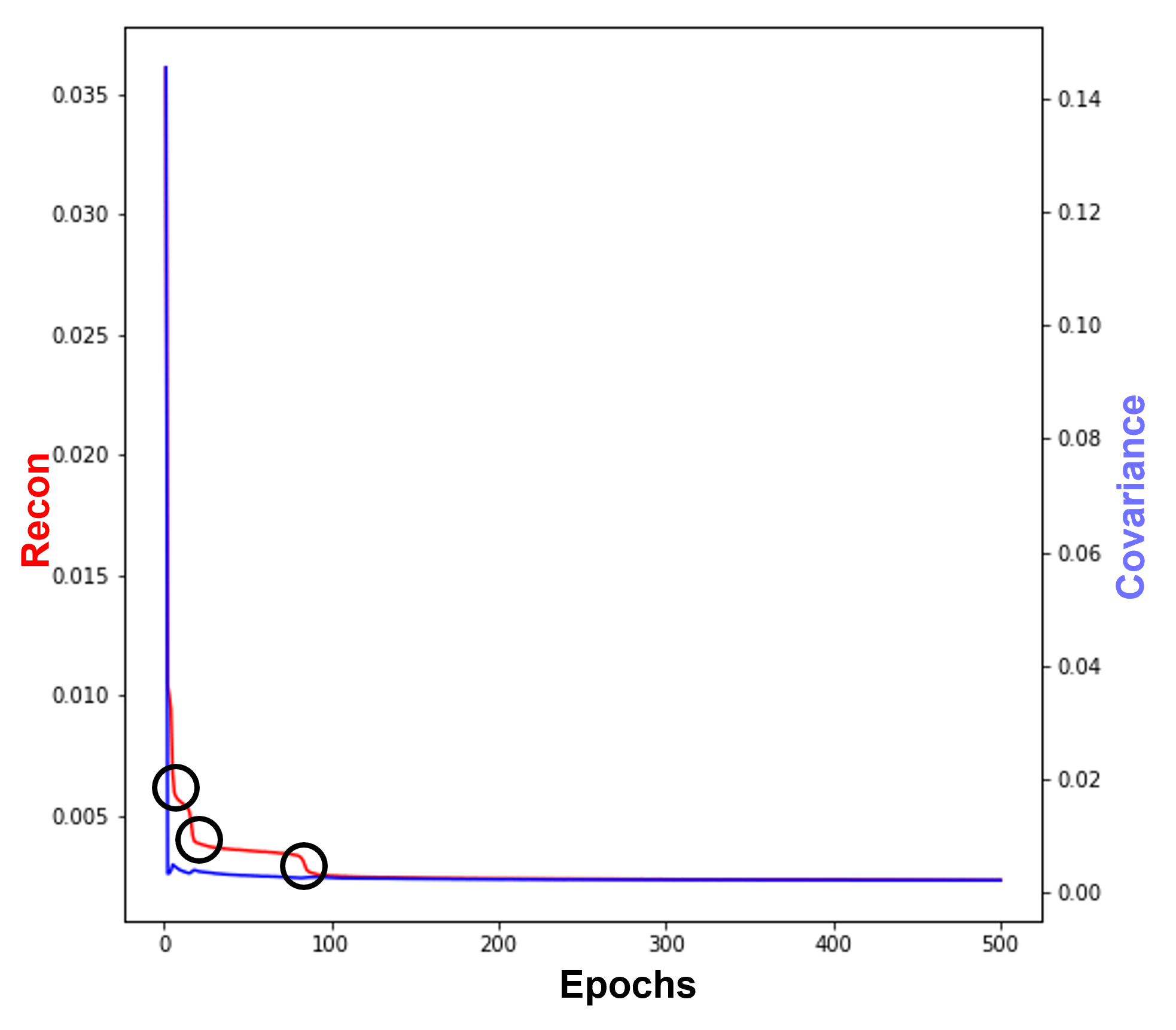}}%
  \caption{Training of Different Variational Auto-Encoders}
  \label{fig:modeltraining}
\end{figure}

\begin{figure*}
\centering
\caption{Encoded Latent Space for STOXX50}
\label{fig:vae_encoding}
\includegraphics[width=0.8\textwidth]{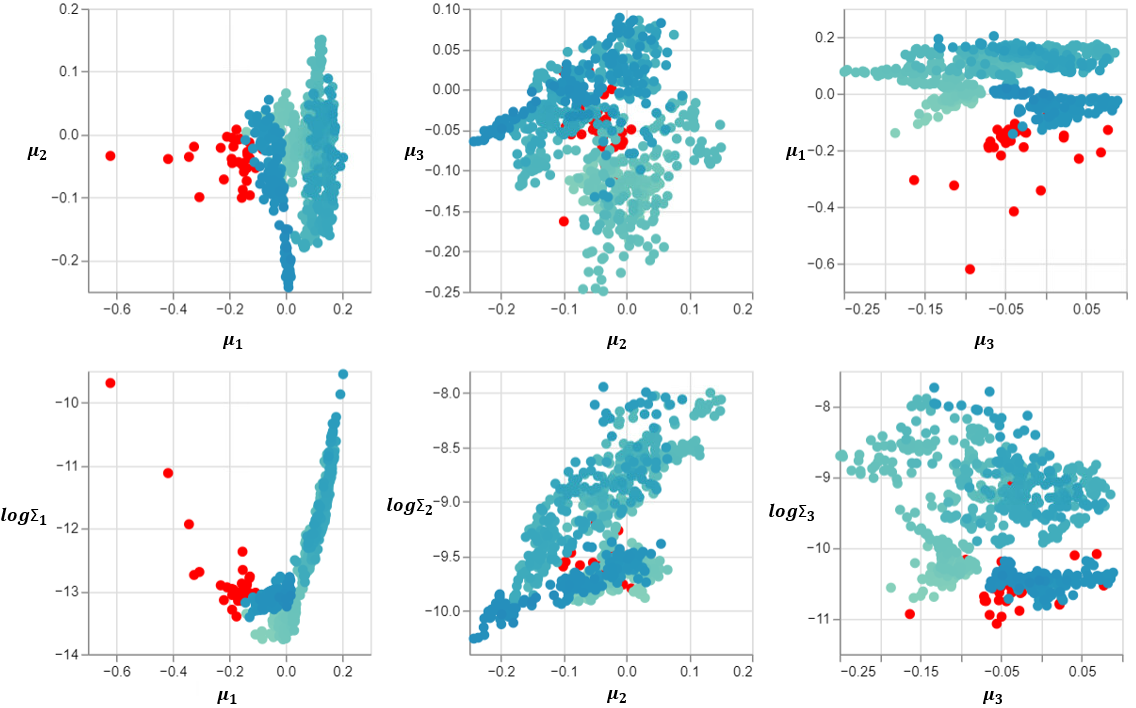}
\end{figure*}

\section{Encoding Implied Volatility Surface}
\label{sec:encoding}
In this section, we are going to compare the training process of a PCA variational auto-encoder with a classic variational auto-encoder model, and discuss our discoveries from the trajectories of the training losses. Once the model is trained, we can obtain the plots for the encoded values of each implied volatility surfaces. As our model is a PCA variational auto-encoder in which input data are encoded into distributions, we are going to plot the mean $\mu$ and logarithm of the variance $\Sigma$. 


\subsection{Training the PCA Variational Auto-encoder}
\label{ssec:training-PCA-VAE}
As mentioned in Section \ref{sec:review}, the PCA variational auto-encoder model incorporates covariance measurements for each pair of latent dimensions in the loss function. By restricting the value of covariance during model training, independent features of volatility surfaces are embedded into latent dimensions. This modification not only largely improves practicability and interpretability of auto-encoder for surface related applications, but also significantly changes the trajectory for the reconstruction losses during model training. 

In this section, we focus on the model training process. It is undoubtedly important to understand what is exactly learnt by the neural network during the training, and this will be a solid foundation for various volatility surface generation applications to be discussed in Section \ref{sec:generate}.

There are three plots in Figure \ref{fig:modeltraining}, which represent the evolution of reconstruction loss $L_{recon}$ in \eqref{eq:pca} and covariance loss $L_{cov}$ in \eqref{eq:lcov} during the training of different variational auto-encoder models. The plots are for a classic variational auto-encoder model, a PCA variational auto-encoder model with three latent dimensions and a PCA variational auto-encoder with four latent dimensions, respectively. The $x$-axis contains the number of training epochs, whereas the $y$-axis measures the loss values. 

Figure \ref{fig:classicialtraining} shows the training process of a classic variational auto-encoder model ($i.e., \lambda_{cov} = 0$) for implied volatility surfaces. We can clearly observe three turning points where the reconstruction loss is significantly reduced; they are at epoch number 2, 7 and 108. Every time the reconstruction loss has a large reduction, the covariance loss is reduced as well, but it quickly grows afterwards. The final value of covariance loss is roughly 0.468.

The training process of a PCA variational auto-encoder (with $\lambda_{cov} = 0.1$) is shown in Figure \ref{fig:PCAtraining}. It is evident that by adding a covariance constraint to the loss function, the training epochs needed to reach the minimum reconstruction error is smaller (reduced from 108 to 35 epochs). The explanation is that since the covariance is limited, the information encoded in the first two latent dimensions are largely reduced so that there is more information to be captured by the third dimension, and the model can quickly learn the feature. On the contrary, for the case of the classic variational auto-encoder, there is little information left for the third dimension, so the model struggles to learn the marginal information. It is important to note that in Figure \ref{fig:PCAtraining} we show then training until 130 epochs for comparison with Figure \ref{fig:classicialtraining}. The actual model we use for our experiments is trained after 40 epochs to avoid over-fitting.

Bergron et al. argue that implied volatility surfaces can be captured by variational auto-encoder using as few as two latent dimensions \cite{bergeron2022variational}. From our experiments of training PCA variational auto-encoder with 2 latent dimensions, we found that even though the reconstruction (MAE) error is small after two dimensions are learnt, the covariance loss is not small enough. More importantly, we did not observe the independence of the latent numbers when we build the interactive generation tool, which will be presented in Section \ref{sec:scenario-generation}.  

We want to explore further the training process by adding one more latent dimension into the PCA variational auto-encoder model; the training process is shown in Figure \ref{fig:PCAtraininglarge}. We keep training the model until 500 epochs and cannot observe any further significant reduction in the reconstruction loss. This indicates that there is no further independent feature to be modelled by the PCA variational auto-encoder model.


This empirical observation can be quantified. It is important to have a quantitative measure, because the number of latent dimensions in the VAE is a hyperparameter that requires calibration and depends on training set and applications. This measure should confirm whether the chosen number of dimensions is too large for a given dataset or if the training process has failed to discover all the expected dimensions. During training it can be used as a stop trigger to end the process after it discovers all the dimensions.

The method we've chosen is to calculate the standard deviation of the encoded training set for each dimension in the latent space and check if they have the same order of magnitude.

The table \ref{tab:z_std} shows the difference between training the PCA variational auto-encoder with 4, 3 and 2 Latent Dimensions. It can be observed that the $Z_{3}$, the additional dimension that failed to capture any new information, has an order of magnitude lower standard deviation.

\begin{table}
\centering
\caption{Standard deviation of encoded training set over variable number of Latent Dimensions for PCA variational auto-encoder}
\label{tab:z_std}
\begin{tabular}{cccc}
\toprule 
& 4 Dimensions & 3 Dimensions & 2 Dimensions \\
$Z_{0}$ & 0.0955 & 0.1172 & 0.1931 \\
$Z_{1}$ & 0.0719 & 0.0957 & n/a \\
$Z_{2}$ & 0.0852 & 0.0997 & 0.1329 \\ 
$Z_{3}$ & 0.0028 & n/a & n/a \\ 
\bottomrule
\end{tabular}
\end{table}


From these observations, we can confirm that three latent dimensions are the only reasonable choice for encoding implied volatility surfaces with PCA variational auto-encoder models. Fewer dimensions will not promote independent features encoded latent dimensions, and adding further dimensions will not encode any extra meaningful features; this will also be emphasised in Section \ref{sec:scenario-generation}.

\begin{figure*}
\centering
\caption{Generated Synthetic Surfaces}
\label{fig:synthetic}
\includegraphics[width=\textwidth]{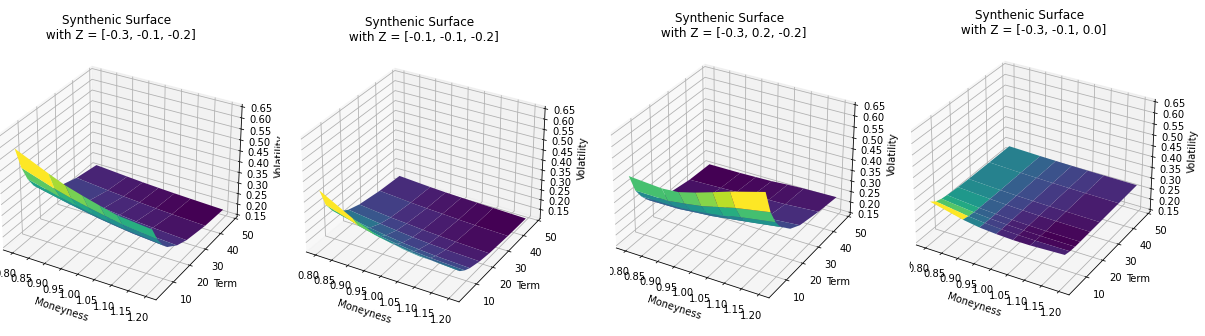}
\end{figure*}

\subsection{Encoded Latent Space}
After a PCA variational auto-encoder is trained, we can generate the latent encoding of our training data and examine how the model transfers the implied volatility surfaces into the latent space. 

We plot the values of $\mu$ and $log \Sigma$ for all the STOXX50 index implied volatility surfaces, see Figure \ref{fig:vae_encoding}. We marked all the surfaces for the trading days in March 2020 and April 2020 as stress scenarios and coloured them in red, as the COVID-19 pandemic caused much turbulence in the market during that period.

In Figure \ref{fig:vae_encoding}, we first observe from the top graphs that $\mu_{1}$, $\mu_{2}$ and $\mu_{3}$ are independent to each other, which is the unique characteristic for PCA variational auto-encoders. If the model is a classic variational auto-encoder, there will be some degree of linear relationship between them. We also discover that the stress scenarios for March and April 2020 are caused by abnormally small values of $\mu_{1}$, which also means that $Z_{1}$ is small, as $Z_{1}$ is sampled from a normal distribution of mean $\mu_{1}$. The stress scenario is not much related to the second or third latent dimensions.

In addition, from the bottom left graph in Figure \ref{fig:vae_encoding}, we found that when $\mu_{1}$ is approximately -0.1, $\Sigma_{1}$ is minimum. As $\mu_{1}$ moves away from -0.1, $\Sigma_{1}$ increases rapidly. From subsequent analysis, which will be discussed in Section \ref{sec:scenario-generation}, we know that the first dimension controls the overall volatility level of a surface. It makes sense that extreme cases are more volatile in the market.

\section{Generating Synthetic Surface}
\label{sec:generate}
We are going to show several important applications of our PCA variational auto-encoder model. The importance and difficulties of these tasks have already been discussed in Section \ref{sec:intro}. We will compare our results with \cite{bergeron2022variational} whenever applicable.

\subsection{Scenario-Based Generation}
\label{sec:scenario-generation}
We mentioned several times that PCA variation auto-encoder is able to encode difference features of implied volatility surfaces independently into three latent numbers. We are going to present how visually this is going to influence the shape of generated synthetic surfaces. There are four synthetic volatility surfaces in Figure \ref{fig:synthetic}, together with the values of latent vector $Z$ which is used to generate each of these surfaces. The first surface on the left is a reference surface, whereas others are generated by changing only one latent values of the reference surface. 

From Figure \ref{fig:synthetic}, we can discover that $Z_{1}$ only controls the overall volatility level, $Z_{2}$ only changes the skewness and $Z_{3}$ affects only for the term structure of a volatility surface. Therefore, we have a powerful method to generate synthetic surfaces for any specific description of the scenario that is requested. For example, if we need a surface of overall high volatility level, small difference of volatility among moneyness and large spread between short and long term, we can use small values for $Z_{1}$ and $Z_{3}$ and a large value of $Z_{2}$ to generate such a surface. This is not possible with the classical variation auto-encoder model.

\begin{table*}
\caption{Volatility surface extrapolation with L-BFGS method}
\label{tab:classic_and_pca_extrapolation}
	\centering
		\begin{subfigure}[b]{0.11\textwidth}
		\centering
	    \begin{tabularx}{\columnwidth}{Y}
            \toprule
            Stock
            \csvreader[head to column names]{stock_list.csv}{}
            {\\ \Stock}
            \\\bottomrule
        \end{tabularx}
    \end{subfigure}
	\begin{subfigure}[b]{0.42\textwidth}
	    \caption{Classic Variational Auto-Encoder Model}\label{tab:classic_extrapolation}
		\centering
	    \begin{tabularx}{\columnwidth}{YYY}
            \toprule
            MAE (known) & MAE (unknown) & Satisfaction
            \csvreader[head to column names]{classic_extrapolation.csv}{}
            {\\ \KP & \UP & \Sat}
            \\\bottomrule
        \end{tabularx}
	\end{subfigure}
    \begin{subfigure}[b]{0.45\textwidth}
        \caption{PCA Variational Auto-Encoder Model}\label{tab:pca_extrapolation}
		\centering
		\begin{tabularx}{\columnwidth}{YYY}
            \toprule
            MAE (known) & MAE (unknown) & Satisfaction
            \csvreader[head to column names]{pca_extrapolation.csv}{}
            {\\ \KP & \UP & \Sat}
            \\\bottomrule
        \end{tabularx}
	\end{subfigure}
\end{table*}

\subsection{Implied Volatility Extrapolation}
\label{sec:extrapolation}
We have briefly introduced the volatility extrapolation problem in Section \ref{sec:intro}. The task is to predict (extrapolate) the full implied volatility surface from only a subset of points on the same surface. 

The extrapolation challenge can be interpreted as a two-step problem. The first step is to make sure that the encoding vector calculated from the known subset of a surface contains enough information to extrapolate the unknown points. This depends on what model is used to generate the encoded latent space. The second step is to use an optimisation method to efficiently find the points on the latent space which represent the known subset of a surface.

Bergeron at el. \cite{bergeron2022variational} suggested to use classic variational auto-encoder to encode the latent space and the L-BFGS algorithm to find the optimal latent encoding vector which minimises the difference on the known subset of a implied volatility surface. To to generate the unknown subset of the surface, they provide this vector to the pre-trained decoder neural network. 

In our dataset, each implied volatility surface has eight terms (3, 6, 9, 12, 18, 24, 36, 48 months) and 7 moneyness settings (0.80, 0.90, 0.95, 1.00, 1.05, 1.10, 1.20). For our extrapolation experiments, we assume the known subset of a volatility surface is of short term (month value of 3, 6, 9 and 12) and close to being at-the-money (moneyness value of 0.95, 1.00 and 1.05), which consists of 12 points. The remaining 44 points are to be extrapolated, and we compare the full generated surface with the full true surface to evaluate the performance of the extrapolation methods using the satisfaction criteria introduced in Section \ref{sec:data}.

We found that for the extrapolation task, PCA variational auto-encoder model is a better solution to produce the encoded latent space. In Section \ref{sec:scenario-generation}, we showed that the model captures the correlation between short- and long-term volatility through latent dimension $Z_{3}$, which indicates that the latent vector found from only short-term information also contain the relevant shape information for the long-term. Because volatility surfaces have to satisfy the no-arbitrage conditions, this connection is pretty stable across stocks and time-horizons. This is also the case for $Z_{2}$, which presents the volatility skewness. This is an important advantage of PCA variational auto-encoder, which leads to better extrapolation performances, as shown in Table \ref{tab:classic_and_pca_extrapolation}.

We calculate the mean absolute errors on the known subset of surfaces (12 points) and unknown subset (44 points) separately in Table \ref{tab:classic_and_pca_extrapolation}, as well as the satisfaction rates among all the 56 points on the surfaces. We can observe that even though both MAE errors are higher for PCA variational auto-encoder, the satisfaction rate achieved is 10\% higher. This indicates that PCA variational auto-encoder is encoding the shape of surfaces instead of individual values of volatility. We also found that the differences of MAEs between two subsets are smaller, which indicates that our model is better to infer the long-term volatility from the short-term volatility as discussed above. However, we think L-BFGS is not a perfect method for optimising on the encoded space because of the level of over-fitting; finding better solutions could be the object of further research.

\subsection{Stock Specific Generation}
The third practical question we want to solve is to infer the implied volatility surfaces for a particular stock from the stock index, by modelling their historical relationships on the latent space produced by a PCA variational auto-encoder model. 

\begin{figure}
\centering
\caption{$Z_{1}$ of single stocks and STOXX50}
\label{fig:z1_plot}
\includegraphics[width=\columnwidth]{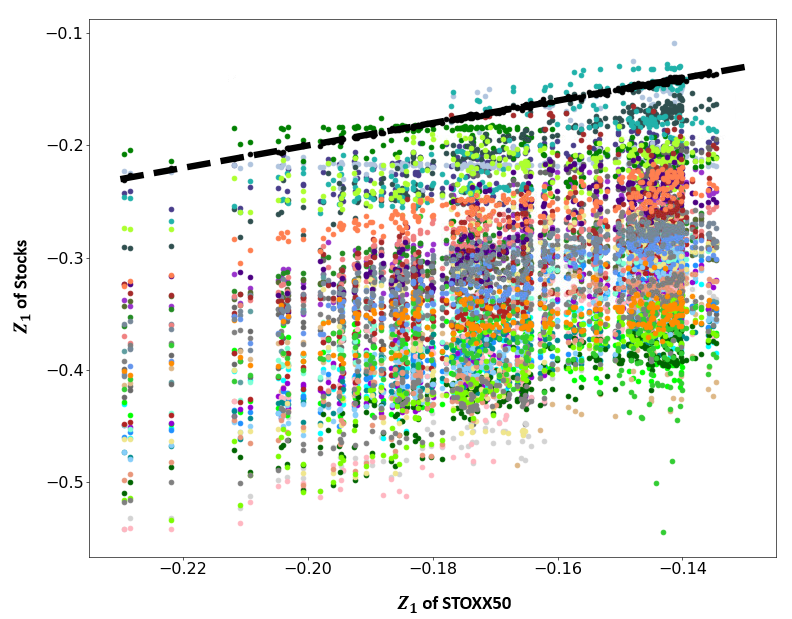}
\end{figure}

For every trading day, we have 39 volatility surfaces, which represent 38 stocks and 1 stock index (STOXX50). We encode these volatility surfaces using a trained PCA variational auto-encoder, and plot for every stock and trading day, its correspondent first latent number ($Z_{1}$) of the stock volatility and STOXX50 volatility surfaces. The plot is shown in Figure \ref{fig:z1_plot}.

Every point plotted in Figure \ref{fig:z1_plot} represents a particular stock on a single trading day. Different colours represent different stocks; we also plot the line $y=x$ as a reference.

We can see From figure \ref{fig:z1_plot} that there is a linear relationship between $Z_{1}$ of each stock and $Z_{1}$ of STOXX50. This relationship exists for all the stocks we plotted in the figure. If we draw lines to models these relationships, we see the lines should have slopes close to one, which indicates if the index volatility surface shifts, the stock volatility surface moves roughly by the same amount. We also notice that most stocks have an intercept smaller than one, meaning that $Z_{1}$ of stocks are generally smaller than $Z_{1}$ of STOXX50. If we refer to volatility surface generation mentioned in Section \ref{sec:scenario-generation}, this should reflect that the volatility surface of a single stock is usually higher than the surface of the STOXX50 index on the same day, a known stylised fact of the market. Similar linear relationships hold true for $Z_{2}$ and $Z_{3}$ as well.

\begin{figure*}
\centering
\caption{Z1 of MUVGn.DE and STOXX50}
\label{fig:muvgn_plot}
\includegraphics[width=\textwidth]{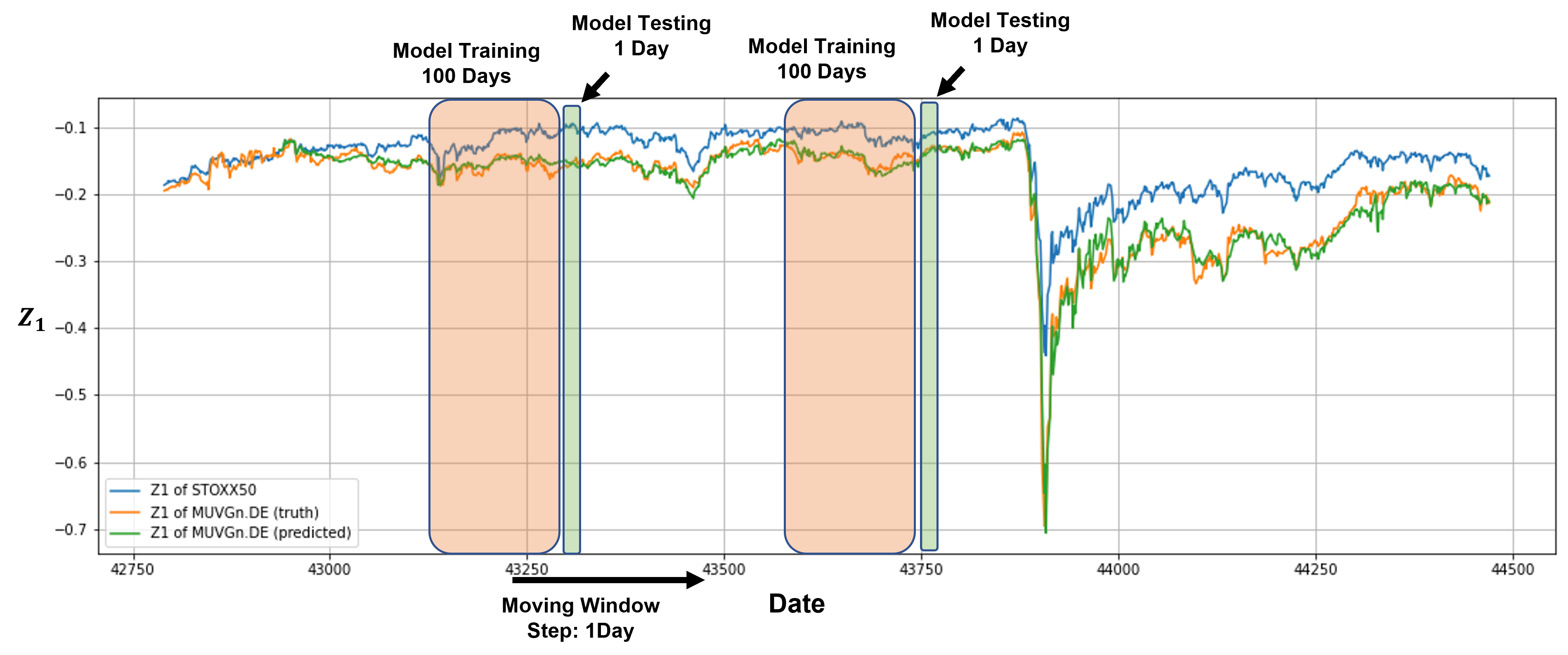}
\end{figure*}

Based on these observations, we use linear regression models to capture the relationship between a particular stock and the STOXX50 index in the encoded space. For every stock on a trading day, we train three linear regression models to predict its three latent numbers separately. The independent variables for linear regression models are the latent number of STOXX50 index and the stock's long-term price volatility. Every linear regression model has two independent variables. The models are trained in a moving windows manner, as illustrated in Figure \ref{fig:muvgn_plot}.


For every trading day in the test period, we use three linear regression models to obtain the latent encoding of the implied volatility surface, which represents , and put the encoding vector into our pre-trained PCA variational auto-encoder. We then compare the generated surface with the true surface in our dataset and obtain the results shown in Table \ref{tab:linear_regression}. The errors in Table \ref{tab:linear_regression} are calculated as the MAE between the true $Z$ value and the predicted one. We see the average satisfaction rate is 76\%, which indicates most predicted implied volatility values are within the bid offer spread of the market. 

\begin{table}[]
\centering
\caption{Predict Stock Volatility Surface Using STOXX50}
\label{tab:linear_regression}
\begin{tabular}{ccccc} 
\toprule Stock & $Z_{0}$ Error & $Z_{1}$ Error & $Z_{2}$ Error & Satisfaction
\csvreader[head to column names]{lr_with_stoxx50.csv}{}
     {\\ \stock & \onemae & \twomae & \threemae & \satisfication} 
     \\\bottomrule
\end{tabular}
\end{table}

\section{Conclusion and Future Research}
\label{sec:conclusion}
We use a new way to encode implied volatility surfaces into a latent space with a PCA variational auto-encoder neural network. The extra covariance measurement in the loss function for model training ensures independence of the three latent dimensions, which brings significant benefits for different synthetic surface generation applications. The scenario-based generation process becomes intuitive and interpretable. The volatility extrapolation performance is largely improved and over-fitting is reduced. We developed a novel solution to infer a single stock volatility surface from the stock index volatility surface.   

Our work can be further developed by finding methods to capture the relationship between a share and an index on the latent space that are better than the linear regression models. It is also possible to use our new encoding for different tasks; e.g., consider the case that there is no historical volatility surface at all for the required stock. We could use a model trained on a difference stock and the stock index to produce the implied volatility surface as required. 

\bibliographystyle{ACM-Reference-Format}
\bibliography{mybib}
\end{document}